\documentclass[iop]{emulateapj}
\usepackage{mathptmx}
\usepackage{natbib}
\usepackage{units}
\usepackage{graphicx}
\usepackage{url}
 \usepackage{amsmath} 	
 \usepackage{color}
\bibpunct{(}{)}{;}{a}{}{,}

\usepackage[colorlinks=true, urlcolor=blue, citecolor=blue, linkcolor=blue]{hyperref}

\newcommand{\be}{\begin{equation}}
\newcommand{\ee}{\end{equation}}

\newcommand{\cii}{\ion{C}{2}}

\newcommand{\oi}{\ion{O}{1}}
	
\newcommand{\mgii}{\ion{Mg}{2}}
\newcommand{\siiv}{\ion{Si}{4}}

\newcommand{\kms}{\ensuremath{\mathrm{km}\,\mathrm{s}^{-1}}}

\newcommand{\ie}{{\it i.e.,}} 
\newcommand{\eg}{{\it e.g.,}} 

\newcommand{\iris}{\textit{IRIS}}
\defcitealias{cii_paper1}{Paper I}
\defcitealias{cii_paper2}{Paper 2}
\def\edt#1{{#1}}

\begin{document}

\title{%
 The formation of  \iris\ diagnostics.
 \\ VIII. \iris\ observations in the \mbox{C II} $133.5$~nm multiplet.
}%
\author{Bhavna Rathore\altaffilmark{1}}
\author{Tiago M. D. Pereira\altaffilmark{1}}
\author{Mats Carlsson\altaffilmark{1}}
\author{ \and Bart De Pontieu\altaffilmark{2,1}}
\affil{
{\altaffilmark{1}Institute of Theoretical Astrophysics, University of Oslo, P.O. Box 1029 Blindern, N-0315 Oslo, Norway}\\
{\altaffilmark{2}Lockheed Martin Solar \& Astrophysics Lab, Org. A021S, Bldg. 252, 3251 Hanover Street Palo Alto, CA 94304 USA}
}
\email{bhavna.rathore@astro.uio.no}
\email{tiago.pereira@astro.uio.no}
\email{mats.carlsson@astro.uio.no}
\email{bdp@lmsal.com}

\begin{abstract}
The \cii\ 133.5~nm multiplet has been observed by NASA's Interface Region Imaging Spectrograph (IRIS) in unprecedented spatial resolution. The aims of this work are to characterize these new observations of the \cii\ lines, place them in context with previous work, and to identify any additional value the \cii\ lines bring when compared with other spectral lines. We make use of wide, long exposure IRIS rasters covering the quiet Sun and an active region. Line properties such as velocity shift and width are extracted from individual spectra and analyzed. The lines have a variety of shapes (mostly single-peak or double-peak), are strongest in active regions and weaker in the quiet Sun. The ratio between the 133.4~nm and 133.5~nm components is always less than 1.8, indicating that their radiation is optically thick in all locations. Maps of the \cii\ line widths are a powerful new diagnostic of chromospheric structures, and their line shifts are a robust velocity diagnostic. Compared with earlier quiet Sun observations, we find similar absolute intensities and mean line widths, but smaller red shifts; this difference can perhaps be attributed to differences in spectral resolution and spatial coverage. The \cii\ intensity maps are somewhat similar to those of transition region lines, but also share some features with chromospheric maps such as those from the \mgii~k line, indicating that they are formed between the upper chromosphere and transition region. \cii\ intensity, width, and velocity maps can therefore be used to gather additional information about the upper chromosphere.


\end{abstract}

\keywords{Sun: atmosphere -- Sun: chromosphere -- Sun: transition region}

\section{Introduction}
The resonance lines from singly ionized carbon around $133.5$~nm are among the strongest lines in the solar ultraviolet spectrum. The multiplet is composed of two main components, at 133.4532~nm (hereafter called the 133.4\,nm line) and at 133.5708~nm which has a weaker blend at 133.5663~nm (hereafter collectively called the 133.5\,nm line). The lines were observed earlier with spectral instruments onboard OSO 8 \citep{1976BAAS....8..501L, 1978ApJ...224..671C} and by the Solar Ultraviolet Measurements of Emitted Radiation \citep[SUMER,][]{1995SoPh..162..189W} instrument onboard the Solar and Heliospheric Observatory (SOHO). Using SUMER observations, \citet{1997ApJ...486L..63C} showed that the \cii\ line intensities are sometimes associated with the intensity in bright grains observed in the UV continuum and neutral lines of C, O and Ni. \citet{1998MmSAI..69..699B} found similar oscillatory behavior in the line intensities observed in the network and internetwork cells. The oscillatory behavior of the \cii\ line shifts was found to be associated with the grains, which are formed deeper in the atmosphere. 

\citet{2003ApJ...597.1158J} compared SUMER observations of the \cii\ lines with 1D radiation hydrodynamic simulations and
concluded that heating by acoustic waves alone was not enough to explain the observations even in the darkest internetwork
areas, thus questioning the notion that the basal flux in stellar chromospheres is caused by non-magnetic heating.

NASA's Interface Region Imaging Spectrograph \citep[\iris,][]{2014SoPh..289.2733D} observes many UV lines in unprecedented spatial and spectral resolution. The \cii\ 133.5~nm multiplet is covered in the FUV window of \iris, and is one of its key observables. \iris\ opens up new possibilities for the study of the structures and dynamics sampled by the \cii\ lines, which before was only possible at much lower spatial resolution.

This is the eighth paper in a series dedicated to the study of the \iris\ diagnostics. With the first papers of the series covering the formation of the \mgii\ lines \citep{2013ApJ...772...89L, 2013ApJ...772...90L, 2013ApJ...778..143P, 2015ApJ...806...14P}, the \cii\ lines have been studied by \citet[][hereafter \citetalias{cii_paper1}]{cii_paper1} and  \citet[][hereafter \citetalias{cii_paper2}]{cii_paper2}. In \citetalias{cii_paper1} the formation properties of the lines were studied using realistic 3D radiative magnetohydrodynamic simulations. It was found that the assumption of optically thin formation breaks down in the majority of cases for the \cii\ lines; their radiation is nearly always optically thick, which has important consequences for the interpretation of observations. The diagnostic potential of the lines was studied in \citetalias{cii_paper2}, where it was shown that the lines have an important diagnostic value, in particular to extract reliable velocities, with a weaker correlation between the line core intensity and temperature also found. It was found that, depending on the profiles of column mass and temperature, the \cii\ lines can be formed both above and below the formation heights of the \mgii~h\&k lines. 

The aims of this work are to provide a general characterization of the \cii\ observations taken with \iris, to discuss their diagnostic value in light of previous work, and to identify how they can complement other chromospheric and transition region diagnostics.   

\section{Observations}
We make use of \iris\ 400-step spectral rasters covering a field of about $140\arcsec \times 180\arcsec$. To maximize the signal to noise ratio (S/N) of the \cii\ lines we used long exposures (30~s). Three datasets were used, two of quiet-Sun and one of an active region. They were used to select five distinct solar regions: network and internetwork quiet-Sun (QS), plage, and sunspot penumbra and umbra. The details of the datasets are given in Table~\ref{tab:obs}. With each raster step taking about $31.6$~s, a full raster took about 3.5~h to complete. All the data were binned on-board by two in wavelength (resulting in a pixel size of 2.56~pm). All but one dataset were unbinned spatially along the slit (having the nominal spatial pixel size of $0\farcs167$).

To complement the \iris\ observations, we also used line-of-sight magnetograms from the Helioseismic and Magnetic Imager \citep[HMI,][]{2012SoPh..275..207S} onboard the Solar Dynamics Observatory \citep{2012SoPh..275....3P}.

\begin{deluxetable}{lllc}
\tablecaption{Observational data sets.\label{tab:obs}}
\tablehead{
\colhead{Target} & \colhead{Starting UT Time} & \colhead{Center coord.} & \colhead{ Binning}  \\
 & & & \colhead{($\lambda$~$\times$~space)}}
\startdata
quiet-Sun & 2014-03-05T11:09:51& ($3\farcs3$, $35\farcs4$) &$2\times$2 \\
Active region & 2014-07-04T11:40:30 & ($-113\farcs3$, $-234\farcs0$) & 2$\times$1\\
quiet-Sun & 2014-07-30T22:55:28&($-327\farcs4$,$-128\farcs1$)&2$\times$1
\enddata
\end{deluxetable}

\begin{figure*}
  \centering
  \includegraphics[width=0.98\textwidth]{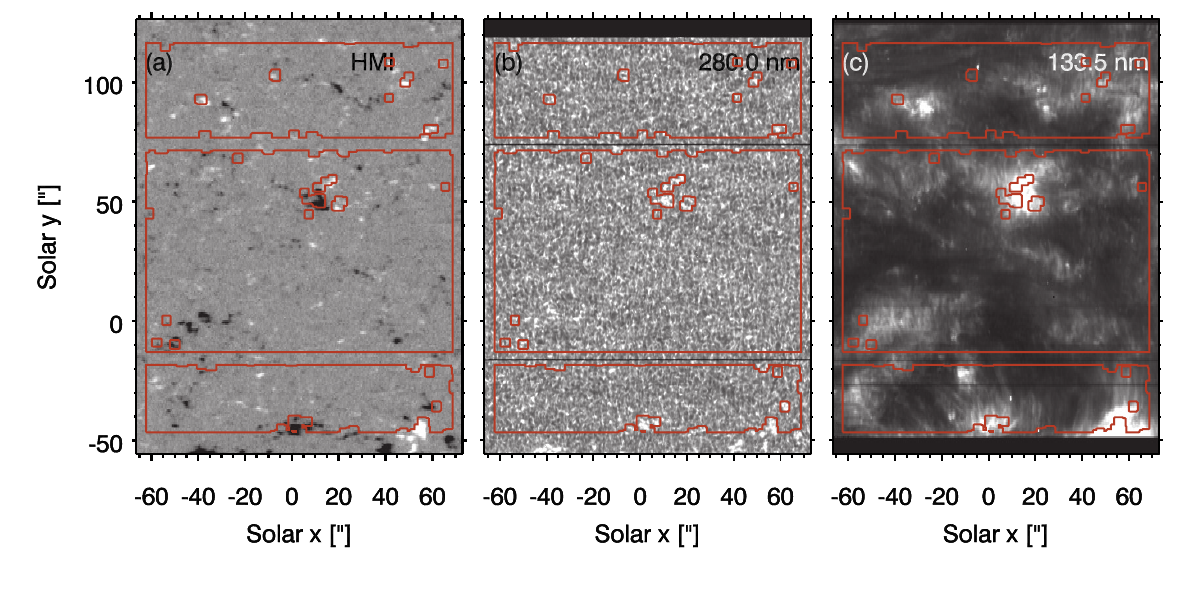} 
  \caption[]{\label{im5m} Quiet-Sun raster from 2014 March 5. Panels depict HMI line-of-sight magnetogram (a) and \iris\ spectral intensities at 280.0~nm (b) and 133.5~nm (c). The red contours represent the region selected as quiet-Sun internetwork. The images are scaled for maximum visibility.
}
\end{figure*}

\begin{figure*}
  \centering
  \includegraphics[width=0.98\textwidth]{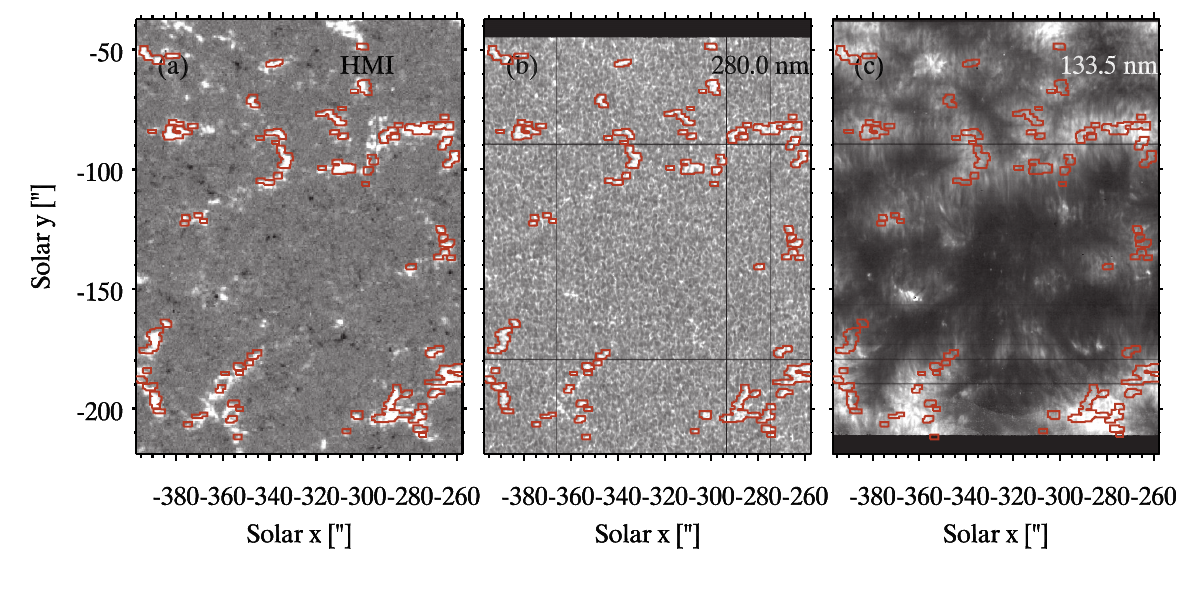} 
  \caption[]{\label{im30j} Quiet-Sun raster from 2014 July 30. Panels depict HMI line-of-sight magnetogram (a) and \iris\ spectral intensities at 280.0~nm (b) and 133.5~nm (c). The red contours represent the region selected as quiet-Sun network. The images are scaled for maximum visibility.
}
\end{figure*}

\begin{figure*}
  \centering
  \includegraphics[width=0.9\textwidth]{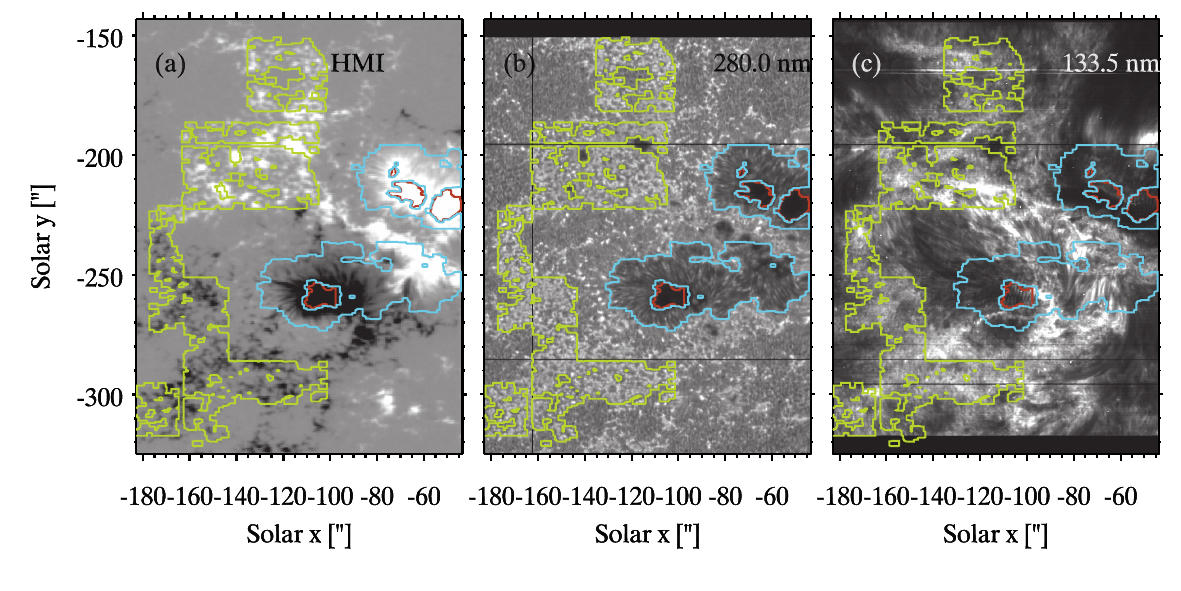} 
  \caption[]{\label{im4j} Active region raster from 2014 July 4. Panels depict HMI line-of-sight magnetogram (a) and \iris\ spectral intensities at 280.0~nm (b) and 133.5~nm (c). The red contours represent the region selected as umbra, the blue contours the penumbra, and the yellow contours the plage. The images are scaled for maximum visibility.
}
\end{figure*}

\subsection{Data Reduction and Region Selection}

We have used \iris\ calibrated level 2 data, details of which are described in \citet{2014SoPh..289.2733D}. To discriminate between different regions we used intensity maps taken at 280.0~nm (quasi-continuum in between the \mgii\ h\&k lines), with different thresholds to select regions such as umbra, penumbra, plage, and quiet-Sun network. The quiet-Sun dataset from 2014 March 5 contains very little network and was therefore selected for sampling the internetwork, while the dataset from 2014 July 30 contains more bright regions and was used for sampling the network. The selected regions are highlighted in Figures~\ref{im5m} (quiet-Sun internetwork), \ref{im30j} (quiet-Sun network), and \ref{im4j} (plage, penumbra, umbra). The two horizontal lines seen in the 280~nm images are \iris\ alignment marks and were ignored in the region selection. 

HMI data were extracted and co-aligned with \iris\ using the following procedure. First, a precise co-alignment between HMI and filtergrams from the Atmospheric Imaging Assembly \citep[AIA, ][]{2012SoPh..275...17L} was obtained by running the \mbox{SolarSoft} routine \texttt{aia\_prep} on both HMI level 1.5 magnetograms and AIA level 1 filtergrams. The \iris\ observations were then co-aligned with HMI and AIA using the \iris\ 140~nm slit-jaw images with the AIA 30.4~nm channel (interpolated to the pixel scale of \iris). The final HMI magnetogram maps were built by extracting the values from the HMI full disk images along the position of the \iris\ slit, using the HMI image closest in time of each \iris\ raster step.

One should keep in mind that each \iris\ raster takes about 3.5~h to complete and the Sun rotates about $32\arcsec$ during this time. With the raster direction being solar east to west, the structures in all maps shown here (e.g. Figures~\ref{im5m}, \ref{im30j}, and \ref{im4j}) appear slightly compressed along the x axis if one were to compare with a similar map taken at one instant in time.

\subsection{Absolute Wavelength and Intensity Calibration}

The \iris\ data used here were processed with version 1.42 of the SolarSoft routine \texttt{iris\_prep}, which includes an improved absolute wavelength calibration that ensures a negligible drift on reference neutral lines such as \oi~135.56~nm, used here as a reference. There are 
also second-order terms included for the wavelength dispersion, which are important for the proper wavelength calibration at 133.5\,nm.
For obtaining the absolute wavelength calibration, we furthermore assume that the spatially-averaged profile along the slit of the \oi~135.56~nm line has a zero velocity shift, so all \cii\ velocity shifts are measured in relation to this line at each slit position.
 
An absolute intensity calibration factor for the FUV window was obtained by using the pre-flight spectrograph effective area (using the SolarSoft routine \texttt{iris\_get\_response}, version 2) and corrected by the sensitivity level as measured by observations of bright stars such as HD91316 (see \emph{IRIS Technical Note 24}).

Earlier OSO 8 observations reported an integrated \cii\ multiplet intensity of 1.13~$\mathrm{W}\;\mathrm{m}^{-2}\;\mathrm{sr}^{-1}$, while from SUMER the total intensities for the multiplet in two internetwork datasets were 0.82 and 0.47~$\mathrm{W}\;\mathrm{m}^{-2}\;\mathrm{sr}^{-1}$ \citep{2003ApJ...597.1158J}. Using \iris\ spectra we measure a \cii\ multiplet integrated intensity of about 0.87~$\mathrm{W}\;\mathrm{m}^{-2}\;\mathrm{sr}^{-1}$ for the internetwork quiet-Sun, which is consistent with earlier measurements.

\subsection{Noise Filtering}
To mitigate the effects of noise in the observed data, we have employed
a weighted mean filter for each individual spectrogram (2D image with
space and wavelength). For the original signal $s$ we took the filtered
signal $s_\mathrm{filt}$ as:

\begin{equation}
 s_\mathrm{filt} =
 \begin{cases}
 \frac{\sigma^2}{\sigma_s^2} m_s + \left(1 -
\frac{\sigma^2}{\sigma_s^2}\right) s  & \text{if } \sigma_s^2 \geq
\sigma^2,\\
  m_s & \text{if } \sigma_s^2 < \sigma^2,
 \end{cases}
\end{equation}
where $m_s$ and $\sigma_s^2$ are the local mean and variance (from a
$3\times 3$ window) and $\sigma^2$ is an estimate of the noise power,
which we took as the average of the local variances. The filtered intensities approach the original signal in regions with high S/N, and tend to the local mean for regions with low S/N (which are typically in the internetwork). 

\edt{The noise filtering increases the number of good fits in the parameter estimations by about 10\% in the internetwork dataset and decreases
the number of outliers. We note that the filtering does not change the overall trend of the results.}

\section{Results}

To study the behavior of the \cii\ 1335\,nm multiplet under various solar conditions, we analyze observations taken in different magnetic field topologies.
An overview of the datasets used is given in Figures~\ref{im5m} (quiet-Sun internetwork), \ref{im30j} (quiet-Sun network), and \ref{im4j} (umbra, penumbra, plage). In the left panels we show the HMI line-of-sight magnetograms, and \iris\ spectra at 280~nm (quasi-continuum in between the h and k lines) and 133.5~nm (\cii\ line core) in the center and right panels. The regions of interest are enclosed by various contours. 

The quiet-Sun internetwork region enclosed by the red contour in Figure~\ref{im5m} appears as mostly gray in the HMI image, without many small dark and bright regions in the magnetogram. The 280.0~nm image shows a mix of bright and dark points everywhere except for a few very bright network points excluded from the selection. The \cii\ line intensity image shows enhanced fibril-like structures along these bright points. The structures are brighter close to the concentrated magnetic field points.  

Figure~\ref{im30j} shows another quiet-Sun region with stronger bright network points. The bright and dark regions in the magnetogram are locations of strong magnetic field, which correlate well with the bright intensity points seen at 280.0~nm. The \cii\ line core intensity shows similar enhanced fibril-like structures along these points with a strong magnetic field.

Figure~\ref{im4j} shows an active region with different structures. The HMI image shows the opposite polarities in both the sunspots and plage. The strong magnetic field region in the plage corresponds to the bright intensity points in the 280.0~nm image. The sunspot umbrae are darker and the penumbrae appear as a mix of dark and gray regions in the 280.0~nm image. The \cii\ intensity image looks very different from the photosphere, representing the upper chromosphere and transition region. The wavelike structures clearly visible at the sunspot umbrae in \cii\ are caused by magneto-acoustic oscillations. Coherent oscillations are seen as vertical structures because of the long raster step cadence (31.7~s).

\subsection{Spatially Averaged Spectra}

\begin{figure}
  \centering
  \includegraphics[width=\columnwidth]{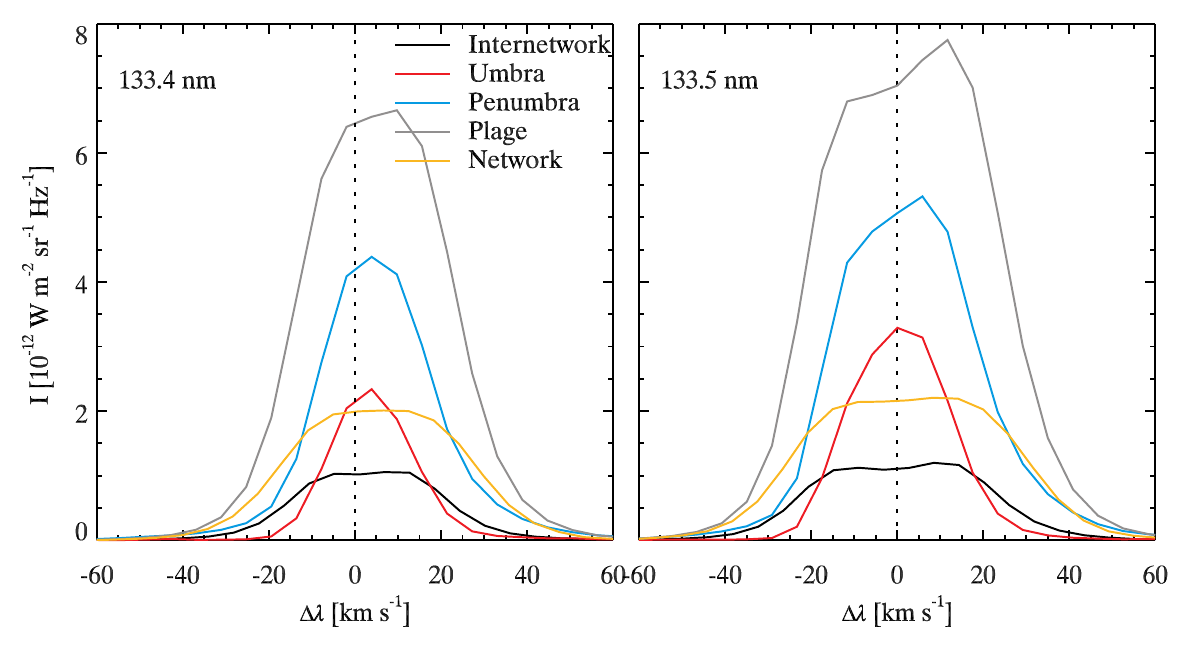} 
  \caption[]{\label{mint}
Mean intensity profiles of the \cii\ 133.4~nm and 133.5~nm lines for different regions. The vertical dotted line shows zero shift with respect to laboratory wavelength. The different curves represent the quiet-Sun internetwork (black), quiet-Sun network (yellow), penumbra (light blue), umbra (red) and plage (gray). 
}
\end{figure}

We have calculated the average spectra for each different region. The average \cii\ line profiles are shown in Figure~\ref{mint}. The line intensity increases from quiet-Sun (QS) internetwork, network, penumbra to umbra, and it is highest in plage. The profiles show a velocity shift with respect to the position at laboratory wavelength (dotted line). Here, positive is red shift. This red shift is smallest in quiet-Sun internetwork. The mean intensity profiles for the internetwork and network regions show a slight central reversal (double-peak profile) in both lines whereas, the umbral, penumbral and plage profiles are single-peak in the 133.4 nm line. The wavelength binning smoothens out the narrow double-peak profiles into single-peak profiles and this affects the number of peak statistics and shape of the mean intensity profiles.

\begin{figure}
  \centering
  \includegraphics[width=\columnwidth]{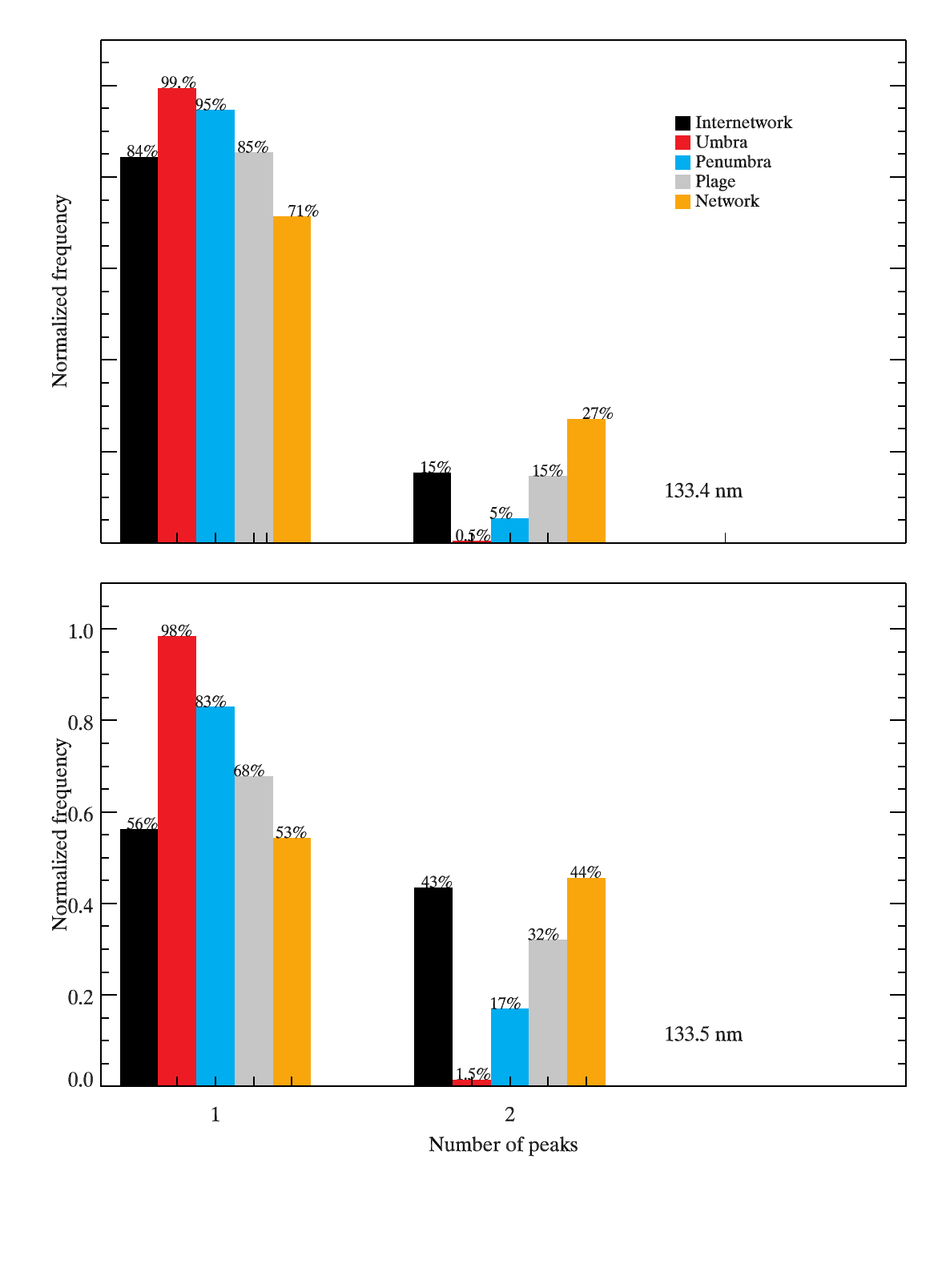} 
  \caption[]{\label{npeaks}
Number of peaks for each \cii\ line in different regions. The percentages for each bin are given at the top of each bar. The bars are colored by region (see legend).
}
\end{figure}

The statistics for the number of peaks in each line in different regions are shown in Figure~\ref{npeaks}. The 133.4~nm line has a larger proportion of single-peak profiles than the 133.5~nm line, which includes a blend. The 133.5~nm line shows about as many single-peak as double-peak profiles in both internetwork and network regions. There were some triple-peak intensity profiles as well, but their number was negligible and they were ignored.

The umbrae of sunspots are a special case where the \cii\ lines (as well as the \siiv\ 139.4~nm and \mgii\ h\&k lines) are mostly single-peak. 

Acoustic waves with periods of 2-3 minutes have been identified in sunspots for decades \citep{1992ASIC..375..261L, 1992ASIC..375....3T}. The umbral oscillations are barely visible in the \siiv\ lines, whereas the \cii\ and \mgii\ line intensities show very prominent oscillations. 
Given the 31.6\,s raster step cadence, these umbral oscillations appear as bright/dark, mostly vertical stripes in the \cii\ intensity in Figure~\ref {im4j}. These wave-like structures belong to magneto-acoustic waves generated below the photosphere, propagating upwards along the field, shocking in the chromosphere and creating umbral flashes.

Such flashes are similar in nature to the internetwork acoustic waves and shocks discussed by \citet{1995ApJ...440L..29C, 1997ApJ...481..500C}, and also to plage shocks and dynamic fibrils \citep{2006ApJ...647L..73H,2007ApJ...655..624D}. These internetwork oscillations can also be seen in the intensities of \mgii\ k and sometimes in \cii\ but they are even more clearly visible as vertical stripes in Doppler shift maps.

\subsection{Line Ratio}

Coronal and transition region lines are typically formed under optically thin conditions, displaying single-peak Gaussian profiles. In these conditions, line ratios from some multiplets can be used to extract atmospheric properties such as density and temperature. However, this is normally not the case with lines that are optically thick. In such cases the line ratio depends on the source functions, hence on both temperature and opacities. In \citetalias{cii_paper1} we studied the intensity ratio between the \cii\ lines and found that when the lines are optically thin, the ratio of the 133.5~nm line to the 133.4~nm line is 1.8 across the whole intensity profile, \ie\ at all frequencies. For optically thick conditions the ratio depends on how the source function varies with height \emph{and} on the ratio between the source functions. 
Double-peak profiles have a source function with a local maximum between the height where the continuum is formed and where the line core is formed. For a case
where the source functions are equal for the two lines, this would lead to equal intensity peaks, a 133\,nm line that is broader and with a deeper self-reversal than 
the 133.4\,nm line and a low total intensity ratio. Single-peak profiles will have a larger total intensity ratio. However, the 133.4~nm line and the weak blend component share the same upper level, which reduces the source function of the 133.4~nm line via radiative rates, increasing the line ratio. 

\begin{figure}[hbtp]
  \centering
  \includegraphics[width=\columnwidth]{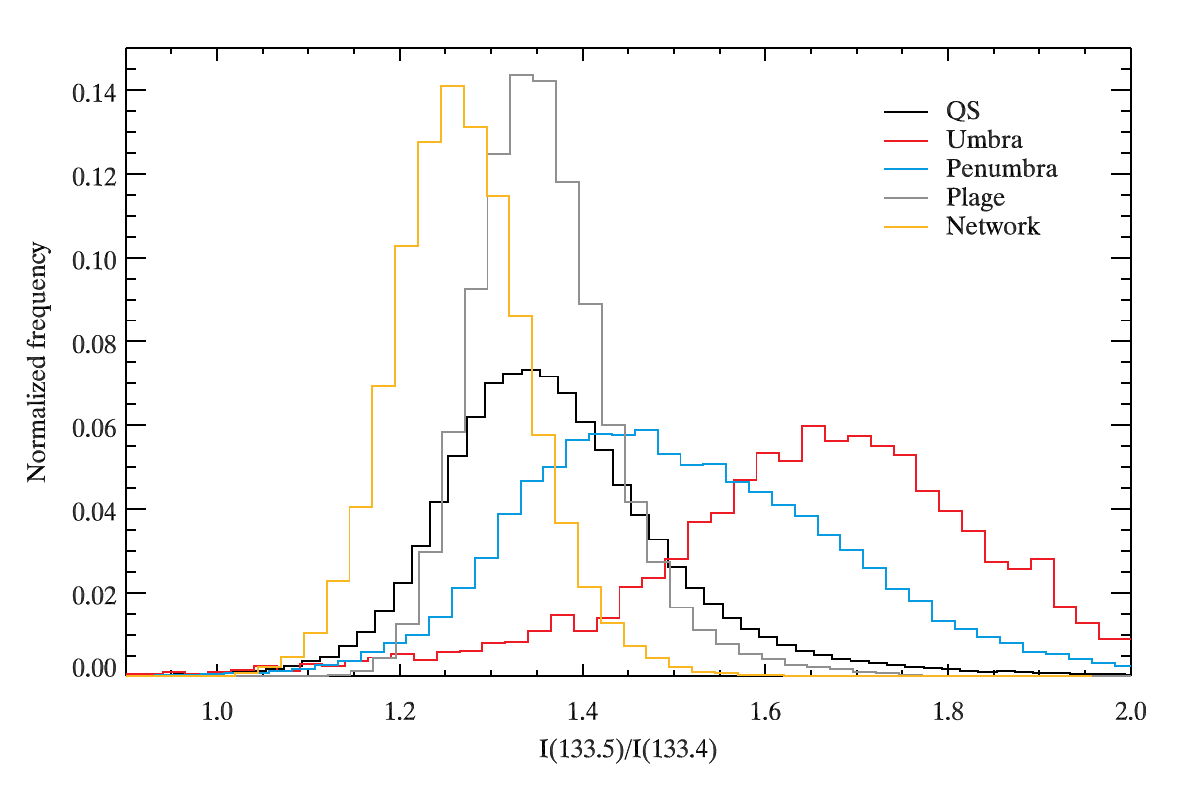} 
  \caption[]{\label{ratio}
Histograms of the ratio of total intensity between the 133.5~nm and 133.4~nm lines, for different regions. The regions are indicated by different colored lines shown in legend. The different curves represent the quiet-Sun internetwork (black), quiet-Sun network (yellow), penumbra (light blue), umbra (red) and plage (gray). 
}
\end{figure}

In Figure~\ref{ratio} we show histograms of the ratio of the total line intensity for the different regions considered.  The line ratio is nearly always less than 1.8, meaning that the lines 
are not dominated by optically thin formation. The network regions show the lowest ratios, peaking around 1.25; the internetwork regions also have
low ratios, peaking around 1.35. Both of these cases of low line ratios can be explained by the large proportion of double-peak profiles in these regions 
(Figure~\ref{npeaks}). The plage region also has ratios similar to the internetwork, although there are more single-peak profiles there. We interpret this as being caused by large densities where the plage profile maxima form, leading to larger collisional rates and therefore similar source functions in both lines.
For the penumbral and umbral regions the proportion of single-peak profiles increases, and so do the line ratios. 

\subsection{Line Peak Intensities, Shifts, and Widths}
To extract the \cii\ line shifts and widths, we fit Gaussians to the line profiles. Despite several line profiles showing non-Gaussian shapes
(\emph{e.g.} double-peak profiles with a central reversal), this approach was favored as it allowed a reliable estimate of the shifts and widths over
the whole field of view (in particular quiet-Sun regions with lower S/N). Other fitting approaches (\eg\ polynomial fitting to the
wings of the line) were tested, but proved less robust and very sensitive to noisy line profiles. One should keep in mind that the Gaussian fits
lead to an underestimation of the line width in the cases with double-peak profiles.

We have measured the peak (maximum) intensities, line shifts ($\Delta\lambda$, in velocity units) and the $1/e$ width of the Gaussian
fit ($\Delta \nu_D$) for all points in the different regions. The line shifts were measured with respect to the \oi\ 135.56~nm line.

\begin{figure*}
  \centering
  \includegraphics[width=\textwidth]{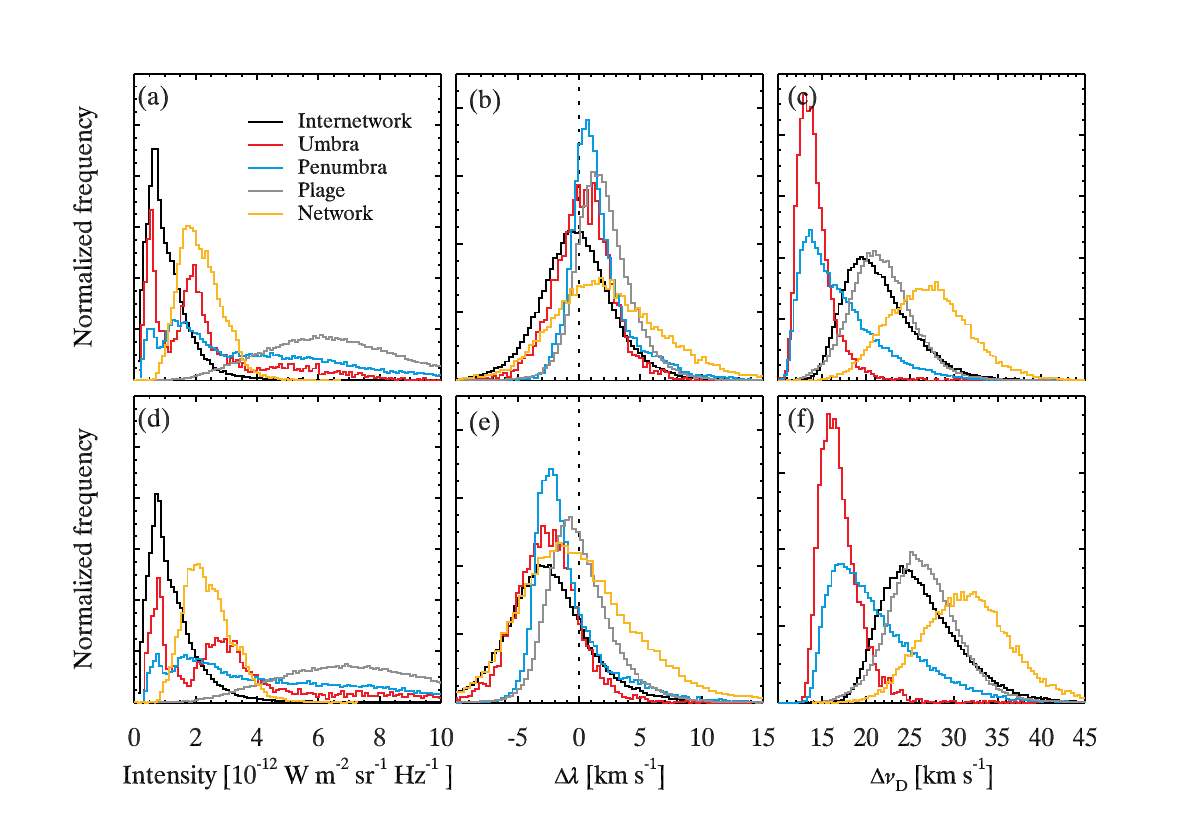} 
  \caption[]{\label{hist_pdw} 
Histograms of extracted quantities in different regions for the \cii\ 133.4~nm (top row) and 133.5~nm (bottom row) lines. Panels depict peak intensity ((a) and (d)), line shift ((b) and (e)) with vertical dashed lines showing zero shift, and line width ((c) and (f)). The different curves represent the quiet-Sun internetwork (black), quiet-Sun network (yellow), penumbra (light blue), umbra (red) and plage (gray).}
\end{figure*}

Histograms of the line properties in different regions are shown in Figure~\ref{hist_pdw}. The peak intensities are highest in plage and smallest in the internetwork. In the umbra the peak intensities have a double peaked distribution, most likely caused by the upward propagating umbral flashes, where the hot shock front is about 1000~K hotter than the surroundings \citep{2013A&A...556A.115D} and increases the \cii\ ionization rates, leading to periodic high intensity spikes in the peak intensities. 

The histograms of line shifts are shown in Figure~\ref{hist_pdw}, panels (b) and (e) for the 133.4\,nm and 133.5\,nm lines, respectively. From the histograms of the line shift one can see that there is approximately a zero shift in the 133.4\,nm line for the internetwork region, whereas the 133.5\,nm line shows a small blueshift of about 2.1~$\kms$ because of the blend. In all the other regions the lines are red shifted with respect to the quiet-Sun internetwork region (black curve). In the plage and network regions the lines are more red shifted compared to the umbra and penumbra (see also Table$~$\ref{t_res} which shows the median of the measured quantities). To avoid the effect of the blend on the shift, we focus on the 133.4\,nm line. The zero shift in the internetwork region and red shifts for all the other regions might indicate that the \cii\ lines show no red shift when there is no magnetic connection to the corona. Also, it is interesting to note that the network region shows a broad distribution of velocity shifts, \ie \ large range of velocities. This may suggest that the flows do not scale with the magnetic field strength, while the lack of mixed polarities 
or the stiffness of the field in plage may cause lower wave amplitudes (and thus a smaller line-of-sight motions). 

The histograms of line widths are given in Figure$~$\ref{hist_pdw}, panels (c) and (f).  Surprisingly,  the plage and the quiet-Sun internetwork show very similar width distributions. In the network one finds the broadest profiles among all regions, while sunspot umbrae show the smallest widths. The penumbrae show a similar peak in the width distribution as the umbrae, but with a broader distribution extending toward wider profiles. The mean widths of individual profiles for all regions are given in Table$~$\ref{t_res}. 
In \citetalias{cii_paper2}, we discussed in detail the width of an optically thick emission line. There are basically two contributions: opacity broadening, which depends 
on how the source function varies through the atmosphere (and is larger for double-peak profiles than for single-peak profiles) and the broadening of the opacity
profile itself (consisting of a thermal and a non-thermal component). It is difficult to separate the two effects without having an independent measure of one of them.

\begin{table}[ht] 
\caption {Average of measured spectral features from different regions. }
  \centering
  	\begin {tabular}{lrrrrrrrrr}
	\hline\hline
		&  \multicolumn{2}{c}{ Shift, $\Delta \lambda$ }& \multicolumn{2}{c}{ Width, $\Delta \nu_D$ }  & \multicolumn{2}{c}{$I_{\mathrm{max}}$}  \\
                 &    \multicolumn{2}{c}{($\kms$)}     &    \multicolumn{2}{c}{ ($\kms$)  }                 & \multicolumn{2}{c}{($10^{-12}\;\;\mathrm{W}\;\mathrm{m}^{-2}\;\mathrm{sr}^{-1}$)}\\
		\hline 
		 &    133.4 & 133.5&   133.4& 133.5 &  133.4& 133.5 \\
		\hline \hline
		Internetwork & $-0.02$ & $-2.3$ & 21.6 & 26.6 & $1.2$ & $1.4$ \\
		Umbra &  0.5& $-2.5$ & 14.4 & 17.2 & $3.1$& $4.5$ \\
		Penumbra  & 1.8  & $-0.7$ & 17.1 & 21.3 & $4.5$ & $5.6$  \\
		Plage & 2.2 & 0.2 & 21.9 & 26.8 & $7.1$ & $ 8.1 $ \\
		Network  &  2.5 & 0.19 & 27.3 & 31.0 & $2.2$& $ 2.4 $ \\
	\hline
	\end{tabular}	

		\label{t_res}
		 \end{table}

\begin{figure*}
  \centering
  \includegraphics[width=\textwidth]{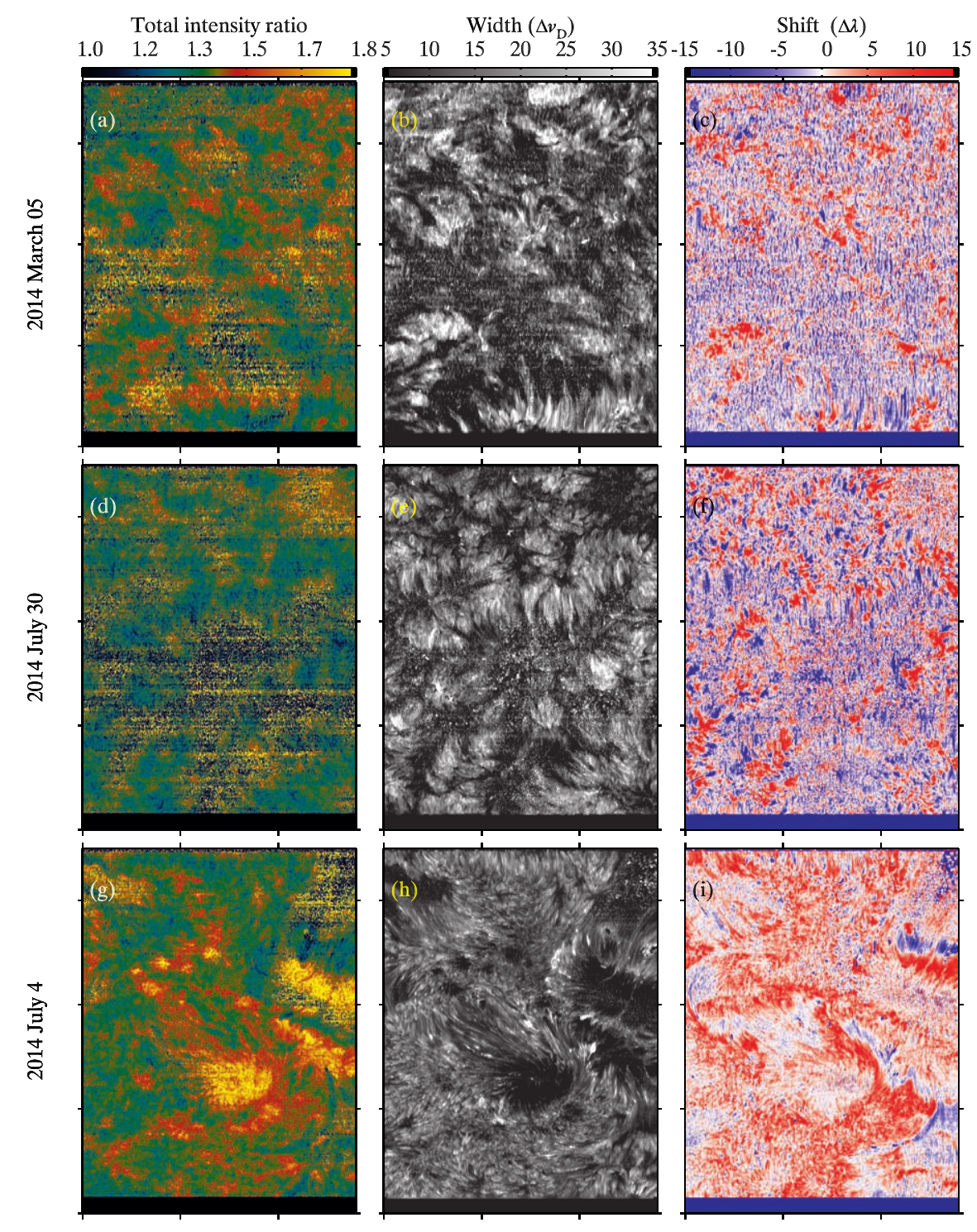} 
  \caption[]{\label{rwd}
  The total intensity ratio of 133.5~nm to 133.4~nm (first column), line width of 133.4~nm (second column) and line shift  of 133.4 nm(third column). The images are scaled to the maximum visibility. The separation between tick marks is 50 arcsec.
  }
\end{figure*}

We find an inverse correlation between the line widths and the line ratios, which can be seen in the maps of Figure~\ref{rwd}.  The structures showing larger line widths (bright structures in panels (b), (e) and (h)) have small intensity ratios, whereas the structures with smaller widths have larger intensity ratios. In \citetalias{cii_paper1}, we showed that double-peak profiles have more opacity broadening than single-peak profiles, and therefore a larger width. They also tend to have a lower intensity ratio. 

The line ratio in the network is typically lower than in other regions (Figure$~$\ref{ratio}), while the widths are largest (Figure~\ref{hist_pdw}, panels (c) and (f)). This suggests that the increased width is partly caused by the 
larger opacity broadening for the double-peak profiles often found in the network. However, this cannot be the whole explanation since the
internetwork has a similar fraction of double-peak profiles (Figure~\ref{npeaks}). This implies that the intrinsic broadening (thermal and non-thermal) must also be larger in the network.
With the chromospheric fibrils that are the disk conterpart of spicules being prominent features in the chromospheric network, the increased line broadening could also be caused by their motions \citep{2007PASJ...59S.655D, 2007ApJ...655..624D,2006ApJ...647L..73H, 2012ApJ...761..138M, 2012ApJ...752..108S, 2013ApJ...764..164S,2013ApJ...769...44S, 2014ApJ...792L..15P}. 

The plage shows similar widths as in the internetwork. We expect higher opacity broadening for the internetwork profiles because of a larger proportion
of double-peak profiles. Therefore the similarity between the
two distributions indicates that plage has larger non-thermal broadening than the internetwork.

\begin{figure}
  \centering
  \includegraphics[width=\columnwidth]{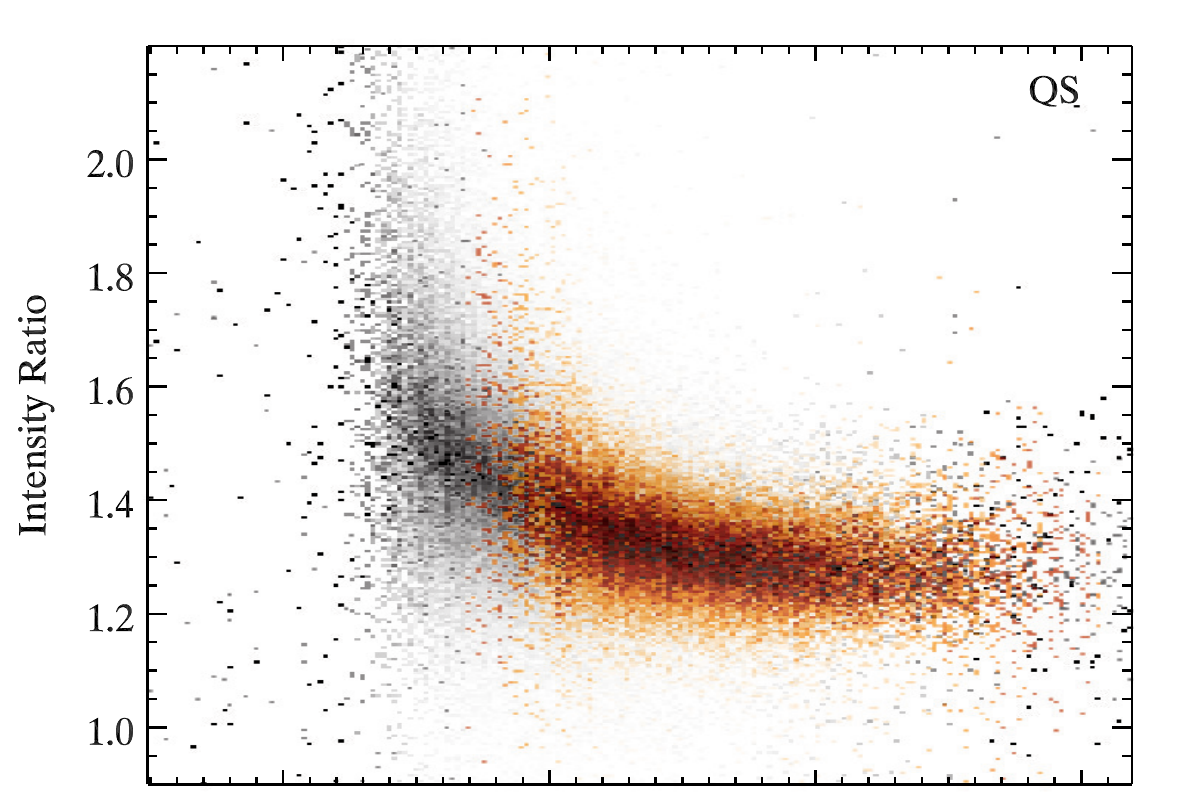} 
  \includegraphics[width=\columnwidth]{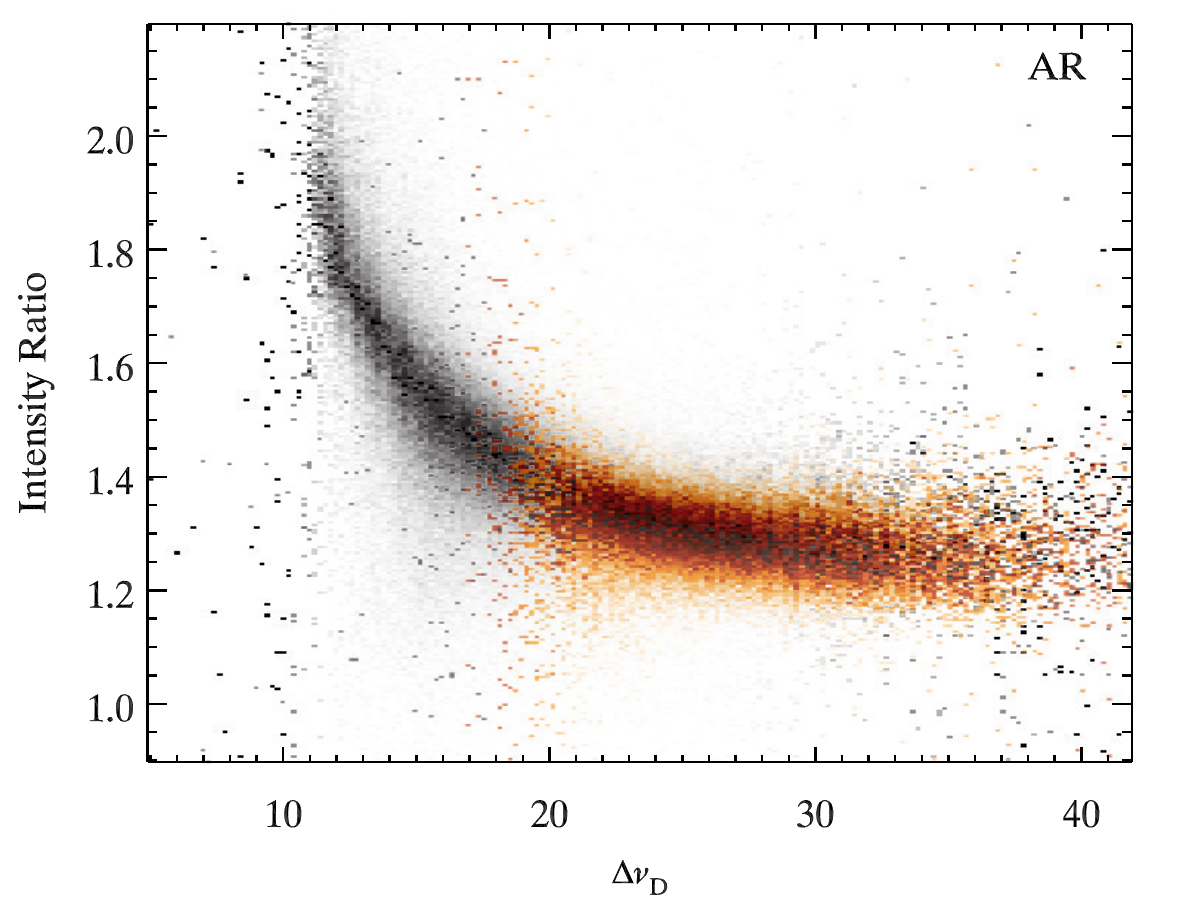} 
  \caption[]{\label{rw_jpdf_qs}
  Probability Density Function of the total line intensity ratio as a function of line width for the quiet-Sun (top panel) and active region (bottom panel), from single-peak (black) and double-peak (red) line profiles.}
\end{figure}

We compared the line width and the ratio of the total line intensity for two data sets, the quiet-Sun (from 2014 March 5) and the active region (2014 July 4). 
In Figure$~$\ref {rw_jpdf_qs} we show the probability density function (PDF) of $\Delta \nu_D$ as a function of the ratio of the total line intensities in the quiet-Sun and the active region. We plot separate distributions for regions with single-peak (black) and double-peak (red) profiles. These PDFs show single-peak profiles typically having smaller widths and larger ratios, whereas the double-peak profiles have typically larger widths and smaller line ratios, consistent with the discussions above.
Note that these panels include all the spectra in each raster, not just the ones selected for the different regions.

\subsection{Comparing \cii \ with \mgii\ and \siiv\ Lines}

The formation of the \cii\ lines has been described in detail in \citetalias{cii_paper2}. From the simulations it was found that the line forms somewhere between the \mgii\ k line core height of formation and the transition region.
However, in the \iris\ observations the lines show many features similar to the \siiv\ 139.4~nm lines (under equilibrium conditions formed around $80\, 000$~K) and sometimes features similar to the \mgii\ k line core intensity. 

\begin{figure*}
  \label{img_3data}
  \centering
  \includegraphics[width=\textwidth]{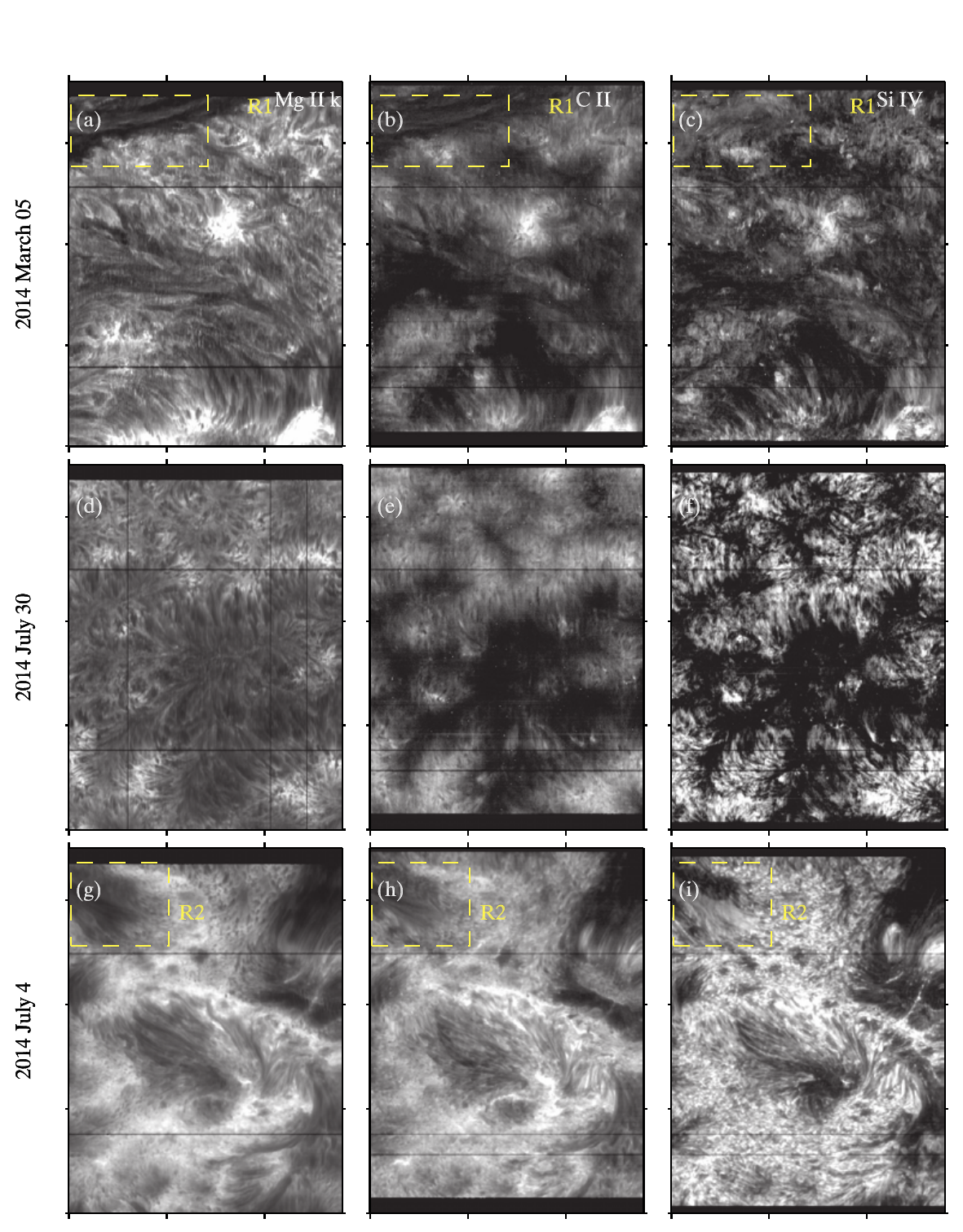} 
  \caption[]{\label{cms_int}
Spectral intensity maps for fixed wavelengths corresponding to the line core of \mgii\ k (left column), \cii\ 133.5~nm (middle column) and \siiv\ 139.4~nm (right column). Images show the different structures in quiet-Sun internetwork (top row), network (middle row) and active region (bottom row). The separation between each tick mark is 50 arcsec along both the axises. The images are scaled for maximum visibility.}
\end{figure*}

In Figure$~$\ref {cms_int} we compare intensity maps taken at the fixed wavelengths of the \mgii\ k$_3$ feature, and the line cores of \cii\ 133.5~nm and \siiv\ 139.4~nm. It is noteworthy to see that in general most of the features in the \cii\ maps bear a strong resemblance to features in both the \mgii\ and \siiv\ intensity maps. On one hand, this could indicate that the structures formed at high temperature, close to the transition region, as seen in \siiv\ are correlated with the upper chromosphere where the \cii\ lines are formed. On the other hand, there are also examples where the \cii\ line maps are very similar to the \mgii\ k maps. Some examples are shown in Figure~\ref {cms_int}, in regions marked $R1$ and $R2$, enclosed in the yellow boxes at the top and bottom rows. In such regions the \cii\ intensity map resembles the \mgii\ k intensity map more than the \siiv\ intensity map.

In the quiet-Sun map, shown in the top row, region $R1$ shows dark structures in both \mgii\ and \cii, whereas these structures are brighter and smaller in the \siiv\ map. These structures resemble filaments, which typically have lower temperatures than mapped by \siiv. Most of the line intensity profiles in region $R1$ are double peaked, with deep central reversals, in both \cii\ and \mgii \ k.

Another area in the active region maps where the \cii\ features resemble both \mgii\ k and \siiv\ is marked as $R2$. This structure is bright in the \siiv \ image (panel (i)), shows mixed bright and dark features in the \cii\ image (panel (h)) and is dark in the \mgii\ k image (panel (g)). We speculate that the dark features in the \cii\ and \mgii\ k intensity maps in this region $R2$ are locations where both the \mgii\ and \cii\ lines are formed higher at lower temperatures, where the source function decreases strongly with height, resulting in the line core intensities being lower for both lines. The brightness in \siiv\, suggests the presence of higher density material at the line's formation temperature in the transition region. The \cii\ Doppler shift maps, especially in the active region (panel (l) in Figure$~$\ref{rwd}) clearly indicate a red shift (downfalling material) in region $R2$. This may be indicating a cool down-flow of coronal material. These downflows are expected to collect more and more material with an increase in depth which can result in larger dark structures in \mgii\ and brighter ones in \siiv.

\section{Discussion and Conclusions}\label{DandC}
\subsection{Intensity}

Both the peak intensity and the profile shape vary between solar regions. The highest intensities are found in plage and the lowest in internetwork 
and umbrae. The intensity profiles may have a single peak or two or more peaks. The number of peaks depends on how the source function varies with height. In the internetwork and network we find mostly double-peak profiles. The network shows slightly more double-peak profiles than the internetwork. In sunspot umbrae the profiles are mostly single-peak. 

The ratio between the total line intensity of the 133.5\,nm and 133.4\,nm lines is almost always lower than 1.8, showing that the formation is not dominated
by optically thin conditions. The ratio is lowest for the regions with the largest proportion of double-peak profiles (network) and highest for the region with predominantly single-peak profiles (umbra).

The comparison between the intensities of the  \mgii\ k, \cii\ 133.5~nm and \siiv\ 139.4~nm lines gives the impression that the \cii\ lines are formed between the formation heights of  \mgii\ k and  \siiv\ 139.4~nm. \cii\ line intensity maps show a mix of features from both the \mgii\ k and the \siiv\ 139.4~nm maps. In most cases, the \cii\ 133.5~nm intensity is similar to that of \siiv\ 140.0~nm. However, some features in the \cii\ maps are similar to \mgii\ k as well.

\subsection{Line Shifts}

The absence of the blend makes the 133.4\,nm line a cleaner diagnostic for velocity shifts. We found its velocity shift to be approximately zero in the internetwork and a red shift of 2.5\,\kms\ in the network. These numbers can be compared with those from SUMER time sequences by \citet{2000A&A...360..742H}. The authors found red shifts of 2.9\,\kms\ and 3.1\,\kms\ in the internetwork and 7.1\,\kms\ and 10\,\kms\ in the network, measured from two different data sets. An estimated error of $\pm 2\;\;\kms$ in the mean velocity was reported for these observations. \citet{1998MmSAI..69..699B}, also using SUMER data, report mean red shifts of 2~$\kms$ and 4~$\kms$ in the internetwork and network, respectively.
We believe that such differences could be explained by the lower spectral resolution of SUMER ($8~ \kms / $pixel vs $2.7 ~\kms /$pixel for \iris), by difficulties in establishing an absolute wavelength scale (given a lack of appropriate neutral lines on both sides of the \cii\ lines), and by the limited spatial sampling of the SUMER observations. These SUMER timeseries observations only sample a $0\farcs3\times120\arcsec$ slit on the Sun, while we have a wide raster sampling about $140\arcsec\times180\arcsec$. Our observations show a large variability, especially between different network patches. Sampling only very few network patches with the SUMER slit may thus lead to differences from our average, which is taken over a large area.

\subsection{Line Widths}

The lines are narrowest in the umbrae of sunspots and successively broader in penumbra, plage, internetwork and broadest in the network. We find average widths
of 22 and 27\,\kms\ for the internetwork and network, respectively. This is comparable to the SUMER measurements from \citet{2000A&A...360..742H}, who find the widths of order 30\,\kms\ (after compensating for the instrumental width) for both internetwork and network. The difference is probably due to the same reasons as
the differences in line shift: a combination of resolution and limited statistics for the SUMER observations.

The narrow profiles found in umbra and penumbra are consistent with the small opacity broadening expected in these regions due to their predominantly single-peak
character (\citetalias{cii_paper1}). The tail in the width distribution to large widths found in penumbrae may be due to a range of non-thermal velocities and a range
of viewing angles to the penumbral filaments.

The network shows the widest profiles. This is partly explained by large opacity broadening expected for double-peak profiles. However, because the internetwork has the same proportion of double-peak profiles, but show narrower profiles, this is only part of the explanation. The network must in addition have large non-thermal broadening of the intrinsic opacity profile. We speculate that this may in part be due to motions in the disk counterpart of spicules that are frequently associated with the network.

Plage shows the same distribution of widths as internetwork. We expect less opacity broadening in plage than in the internetwork because of the smaller number of
double-peak profiles with large peak separation. The equality in the width distribution, thus indicates larger intrinsic broadening (thermal and non-thermal) for plage.

The line widths for the \cii\ lines show an inverse correlation with the intensity ratios. The narrower single-peak intensity profiles are  formed in the lower density upper atmosphere, whereas the broader lines with a central reversal have the peaks formed in the deeper atmosphere at higher densities. The double-peak lines are broader because of a larger opacity broadening factor (see \citetalias{cii_paper1}).

\subsection{Conclusions}

Our results confirm that the \cii\ lines are good chromospheric diagnostics. The number of peaks, intensities, line ratios, shifts, and line widths show
distinct differences between different types of solar regions. The structures observed in \cii\ intensity maps are well correlated with structures seen in \siiv\ maps and in fewer cases with structures seen in \mgii\ k. The \cii\ lines can therefore be used as a tool to link the structures observed with other observables from the lower chromosphere to the transition region. In some cases, they can be distinctly different from other \iris\ lines such as \mgii~h\&k or \siiv~139.4~nm and, \emph{e.g.} through their width maps, can provide a rich amount of information that is unique among the \iris\ diagnostics.

\begin{acknowledgements}
The research leading to these results has received funding from the European Research Council under the
 European Union's Seventh Framework Programme (FP7/2007-2013) / ERC grant agreement no 291058.
This research was supported by the Research Council of Norway through the grant ÒSolar Atmospheric 
ModellingÓ and through grants of computing time from the Programme for Supercomputing and
through computing project s1061 from the High End Computing Division of NASA.  
B.D.P. acknowledges support from NASA grants NNX11AN98G and NNM12AB40P and NASA contract NNG09FA40C (\iris).
\iris\ is a NASA small explorer developed and operated by LMSAL with mission operations executed at NASA Ames and major contributions to downlink communications funded by ESA and the Norwegian Space Centre.
\end{acknowledgements}

\bibliographystyle{apj}   
\bibliography{CII_paper3}        
\end{document}